\documentstyle[aps,preprint,eqsecnum]{revtex}
\tightenlines

\def \be {\begin{equation}}
\def \eq {\end{equation}}
\def \bee {\begin{eqnarray}}
\def \eqq {\end{eqnarray}}

\def \bea {\begin{array}{c}}
\def \eqa {\end{array}}

\def \R {{\bf R}}

\def \del {\partial}
\def \dels {\partial\kern-.5em / \kern.5em}
\def \As {{A\kern-.5em / \kern.5em}}
\def \Ds {D\kern-.7em / \kern.5em}

\def \a {\alpha}
\def \b {\beta}
\def \g {\gamma}
\def \G {\Gamma}

\def \lam {\lambda}

\def \O {\Omega}

\begin{document}
\draft
\title{The $c$-Functions of Noncommutative
Yang-Mills Theory \\
from Holography}
\author{Feng-Li Lin and Yong-Shi Wu}
\address{Department of Physics, University of
Utah, Salt Lake City, UT 84112, U.S.A.}
\date{\today}
\maketitle
\begin{abstract}
In this paper we study non-commutative
Yang-Mills theory (NCYM) through its
gravity dual. First it is shown that
the gravity dual of an NCYM with self-dual
$\theta$-parameters has a Lagrangian in the
form of five-dimensional dilatonic gravity.
Then we use the de-Boer-Verlinde-Verlinde
formalism for holographic renormalization
group flows to calculate the coefficient
functions in the Weyl anomaly of the NCYM
at low energies under the assumption of
potential dominance, and show that the
$C$-theorem holds true in the present case.
\end{abstract}

\newpage

\section{Introduction}
Yang-Mills theory on a non-commutative
space \cite{Conn,CR}, or simply
noncommutative Yang-Mills theory
(NCYM), has recently received
increasing attention in string theory
community. A few years ago,
coordinates for coincident D-branes
were shown to naturally promote to
matrices \cite{Witt0}, signaling the
relevance of noncommutative gauge
theory \cite{HW1}. Later, NCYM was
shown to actually appear in the
D-string solution to the IIB matrix
model \cite{Li0} and in various
string/M(atrix) theory
compactification with constant NS-NS
B-background
\cite{CDS,DH,HWW,HW2,HW3}. This is not
too surprising, because for a single
D-brane in a background with constant
gauge field-strength or rank-two
anti-symmetric B-tensor, some
appropriate limit should lead to a
situation similar to that of a
particle in the lowest Landau level,
where the guiding-center coordinates
are known to be non-commuting. Indeed
in a recent seminal paper, among other
things, Seiberg and Witten \cite{SW}
have explicitly identified the precise
limit in string theory for NCYM to
work, which is similar to the limit in
M theory for discrete light cone
quantization of Matrix theory to work
\cite{SenSei}. In this way, NCYM
arises as a new limit in string
theory, providing a new probe to
non-perturbative effects in
string/M(atrix) theory.

In this paper we study NCYM by
exploring its supergravity dual. By
now it is widely believed that gauge
theory is dual to a certain limit of
string theory \cite{Mald,GKPW}; in
particular, type IIB supergravity on
an anti-de Sitter background, say of
five dimensions, can be used to
describe a large-$N$ supersymmetric
Yang-Mills (SYM) theory on the four
dimensional boundary, which is known
to be a conformal field theory (CFT).
One important test of this AdS/CFT
correspondence is the holographic
derivation of the quantum Weyl anomaly
in the $D=4$, ${\cal N}=4$, $SU(N)$
SYM from the generally covariant boundary
counter-terms in the classical action
of its bulk AdS gravity dual \cite{HS},
with central charges reproducing the
expected large-$N$ behavior.
It seems natural to extend this
correspondence between gauge theory
and gravity to NCYM. The supergravity
backgrounds with non-vanishing
B-fields that are supposed to be dual
to NCYM have been recently suggested
in refs. \cite{HI} and \cite{MR,ASO}.
Furthermore, it was observed in ref.
\cite{LW} that these supergravity
duals can be derived from the
Seiberg-Witten relations \cite{SW}
between closed and open string moduli,
by assuming the running string tension
is a simple power function of energy.
This observation suggests that one
should try to use the supergravity
duals to explore the running behavior
of NCYM.

It is known that NCYM is no longer
conformally invariant, because of the
length scale associated with a
non-vanishing B-background. Thus, one
expects that the "central charges" of
NCYM, defined as the coefficients in
its Weyl anomaly, should run as a
function of the energy scale. It is
interesting to calculate these functions,
the so-called $c$-functions, and to see
whether they obey a generalization of the
famous $C$-theorem \cite{c-thrm} in two
dimensions, that asserts the $c$-function
is always monotonically increasing with
the energy scale. The consistent coupling
of NCYM to a curved background is not
known yet, so it is not possible at this
moment to directly calculate the Weyl
anomaly of NCYM on the field theory side.
The goal of the present paper is to study
the $c$-functions in a holographic manner
through the supergravity dual, thus
providing constraints and shedding light
on the problem of consistently coupling
NCYM to a curved background.

The method we are going to use to
calculate the holographic Weyl anomaly
is the Hamilton-Jacobi approach to the
5-dimensional bulk gravity developed
by de Boer, Verlinde and Verlinde\cite{BVV}.
In this approach an analogue of the
first-order Callan-Symanzik equations for
the 4-dimensional dual field theory on
the boundary can be derived from the
bulk Hamilton-Jacobi equations. The key
point is to interpret the Hamilton-Jacobi
functional as the quantum effective action
in the dual field theory resulting from
integrating out the matter degrees of freedom
coupled to the boundary background gravity.
Then from it one can derive the holographic
Weyl anomaly and $c$-functions. Moreover, in
the de Boer-Verlinde-Verlinde formalism there
are dilaton-like scalar fields in 5-dimensional
gravity. The radial profile of these fields in
the bulk represents the renormalization group
(RG) running of certain coupling constants in
the dual field theory on the boundary. To
apply this formalism, one needs to show that
the NCYM's gravity dual given in \cite{MR}
for the full 10-dimensional IIB background
really has a 5-dimensional dilatonic gravity
Lagrangian after dimensional reduction. In
this paper we will show that this is indeed
the case for an NCYM with self-dual
$\theta$-parameters, corresponding to
isotropic non-commutativity, whose gravity
dual has a self-dual B-background, such
that the B-field does not explicitly appear
in the action for the dilaton-gravity sector
after dimensional reduction to 5 dimensions.

The paper is organized as follows. In Sec.
II we show that after dimensional reduction
from 10-dimensional IIB supergravity, the
gravity dual of an NCYM with self-dual
non-commutativity parameters has a
5-dimensional Lagrangian in the form of
a dilatonic gravity. In Sec. III the
de Boer-Verlinde-Verlinde formalism for
holographic renormalization group flows
is adopted to calculate the $c$-functions
of the NCYM at low energies under
the assumption of potential dominance.
In Sec. IV, we show that the $c$-functions
defined in Sec. II transform as vectors on
the dilaton-space as the beta function does.
The final section, Sec. V, is devoted to
conclusions and discussions. In the appendix
we show how to generalize the
de Boer-Verlinde-Verlinde formalism to
the non-canonical form of dilatonic gravity.

\section{Dilaton Gravity Dual of
Noncommutative Yang-Mills}

In contrast to the gravity dual of
ordinary Yang-Mills theory, the
supergravity dual of NCYM involves
turning on nontrivial scalar and
various $r$-form (anti-symmetric
$r$-tensor) backgrounds with a radial
profile\cite{MR}. One may wonder
if there exists a 5-dimensional
gravity action from which the
equations of motion dimensionally
reduced from 10-dimensional
IIB supergravity can be derived. We
will show that there is indeed such a
5-dimensional dilatonic gravity action,
at least for the case with self-dual
B-backgrounds.

The bosonic action for type IIB
supergravity in ten dimensions in the
Einstein frame is
\be
I_{10}= {1\over 2 \kappa^2_{10}} \int
d^{10}z \sqrt{det\;G}\; [R - {1\over
2} (\partial \phi)^2 -{1\over 2}
e^{2\phi}(\partial \chi)^2 -{1\over 2
\cdot 3!} e^{-\phi} H^2_3 - {1\over 2
\cdot 3!} e^{\phi} F_3^2 -{1\over 4
\cdot 5!} F_5^2 ]\;,
\label{10dsugra}
\eq
where $\phi$ is the dilaton, $\chi$
the RR scalar and the form strengths
are defined as
\bee
H_3&=&dB_2 \;, \\
F_3&=&dA_2 - \chi H_3 \;, \\
F_5&=&dA_4 -{1\over 2} A_2 \wedge H_3
+ {1\over 2} B_2 \wedge F_3 \;.
\eqq
Here $B_2$ and $A_2$ are respectively
the NS-NS and RR 2-form potentials,
$A_4$ are the RR 4-form potential and
the 5-form strength $F_5$ is
self-dual, that is
\be
*F_5=-i\;F_5\;,
\label{selfdual}
\eq
where $*$ is the 10-dimensional Hodge
dual.

In this paper, we only consider
self-dual B-backgrounds, together with
the following conditions motivated by
supersymmetry\cite{Das}:
\bee
\chi&-&i e^{-\phi}=i c \;,
\nonumber\\
F_3&=&i c H_3 \;,
\nonumber\\
B_{0 1}&=&B_{2 3}\;, \qquad A_{0
1}=A_{2 3}\;.
\label{B}
\eqq
where $c$ is a real constant.

These conditions are consistent with
the equations of motion for scalars
and two-form potentials. Moreover,
they make the contributions to the
energy-momentum tensor from the NS-NS
and RR sector cancel each other, except
the one from self-dual five-form
strength. The Einstein equations and
the Gauss law for the five-form strength
thus form a closed set of equations:
\bee
R_{MN}=T_{MN}^F \;, \label{10d}
\\
\partial_{M}(\sqrt{det\;G} F^{MNPQR})=0
\;,
\label{gausslaw}
\eqq
where $T_{MN}^F={1 \over 4 \cdot
4!}F_{MPQRS}F_N^{PQRS}$ is the
energy-momentum tensor of the
five-form strength. Once the
solution to this set is known, the
other fields can be solved from
their equations of motion.

We now perform dimensional
reduction by using the following
ansatz for the 10-dimensional metric:
\be
ds^2=G_{M N}(z) dz^M dz^N=
\hat{g}_{mn}(x) dx^{m} dx^{n} +\ell^2
\Phi(x) \bar{g}_{a b}(y) dy^a dy^b \;,
\label{g10}
\eq
with the indices \{$m,n$\} running on
the reduced 5-dimensional manifold
$\cal{M}$, and indices \{$a,b$\} on a
prescribed 5-sphere with $x$-dependent
radius $\sqrt{\ell^2 \Phi(x)}$, where
$\ell^2$ is the typical length scale
of $\cal{M}$ related to the D3-brane
charge (or the 5-form flux), and the
Jordan-Brans-Dicke scalar $\Phi$ turns
out to be the dilaton in the reduced
theory.

Using this ansatz we can solve the
Gauss law equation (\ref{gausslaw})
for the five-form strength, and the
solution is
\be
F_{mnopq}= i 4 \ell^{-1} \Phi^{-5/2}
\sqrt{det\;\hat{g}}\;\;
\epsilon_{mnopq}\;,
\eq
where the $\epsilon-$symbol is equal to
$1$$(-1)$ for even (odd) permutations of
$0,1,2,3,r$, and to $0$ otherwise.

The components of the corresponding
energy-momentum tensor are
\be
T^F_{mn}=-{4 \over \ell^2} \Phi^{-5}
\hat{g}_{m n}\;, \qquad T^F_{ab}= {4
\over \ell^2} \Phi^{-4} \bar{g}_{a
b}\;, \qquad T^F_{ma}=0 \;.
\eq
The 10-dimensional Einstein equation
(\ref{10d}) then decomposes into
\bee
R_{mn}&=&\hat{R}_{mn} -{5 \over 2}
\Phi^{-1} \hat{\nabla}_{m}
\hat{\nabla}_{n} \Phi +{5 \over 4}
\Phi^{-2} \hat{\nabla}_{m}\Phi
\hat{\nabla}_{n} \Phi = -{4 \over
\ell^2} \Phi^{-5} \hat{g}_{mn}\;,
\label{eom1}
\\
R_{ab}&=&({4 \over \ell^2} -{1 \over
2} \hat{\nabla}^2 \Phi - {3 \over 4}
\Phi^{-1} (\hat{\nabla} \Phi)^2 )
\bar{g}_{ab} = {4 \over \ell^2}
\Phi^{-4} \bar{g}_{ab}\;,
\label{eom2}
\\
R_{ma}&=&{1 \over 2}\Phi^{-1}
\partial_{m} \Phi \bar{g}^{bc} (
\bar{\nabla}_c \bar{g}_{ab} -
\bar{\nabla}_a \bar{g}_{bc})=0\;,
\label{eom3}
\eqq
where $\hat{R}_{mn}$ is the Ricci
tensor of the metric $\hat{g}_{mn}$,
while $\hat{\nabla}$ and $\bar{\nabla}$
the covariant derivative with respect
to $\hat{g}_{mn}$ and $\bar{g}_{ab}$
respectively.

Given the prescribed 5-sphere metric
$\bar{g}_{ab}$, eq. (\ref{eom2})
reduces to a single equation of motion
for the dilaton $\Phi$, and eq.
(\ref{eom3}) is just an identity
because of the metricity condition.

It turns out that this reduced set of
equations of motion (\ref{eom1}) and
(\ref{eom2}) for $\hat{g}_{m n}$ and
$\Phi$ can be derived from the
following 5-dimensional action for a
dilatonic gravity
\be
I_{5}={V_5 \over 2 \kappa^2_{10}}
\int_{\cal{M}} d^{5}x
\sqrt{det\;\hat{g}}\;\Phi^{5/2}[
\hat{R}+5\Phi^{-2}(\hat{\nabla}
\Phi)^2 )+\ell^{-2}(20 \Phi^{-1}- 8
\Phi^{-5})]\;,
\label{noncan}
\eq
where $V_5$ is the volume of the unit
5-sphere. This action reduces to the
familiar action for AdS gravity if we
set $\Phi=1$.

To bring the gravity action to the
canonical Einstein-Hilbert form we
need to do the following Weyl
transformation
\be
\hat{g}_{mn}=\Phi^{-5/3} g_{mn}\;,
\label{weylt}
\eq
and the corresponding new
5-dimensional action is
\be
I^{EH}_{5}={V_5 \over 2 \kappa^2_{10}}
\int_{\cal{M}} d^{5}x
\sqrt{det\;g}\;[R- {10 \over 3}
\Phi^{-2}( \nabla \Phi)^2
+\ell^{-2}\Phi^{-8/3}(20- 8
\Phi^{-4})]\;,
\label{can}
\eq
where the unhatted quantities are with
respect to the new 5-dimensional
metric $g_{mn}$.

This action was derived before from
the gauged supergravity point of
view in a different context\cite{CL}.
Our above discussions establish that
the proposed gravity dual\cite{MR}
of an NCYM with isotropic non-commutativity
has a 5-dimensional dilatonic gravity
description given by the action
(\ref{noncan}) or (\ref{can}).

One may feel odd at first sight that
in the 5-dimensional reduced dilatonic
gravity dual, the 2-form field that
specifies the non-commutativity in the
original boundary Yang-Mills theory does not
show up explicitly. The puzzle is resolved
by noticing that the self-duality conditions
(\ref{B}) place strong restrictions on
the holographic profile of the dilaton
$\Phi$ to make it dependent on the
asymptotic value of the 2-form field.
This is most easily seen from
the full expression of Maldacena and
Russo's solution\cite{MR} (in the near
horizon limit):
\bee
&& ds^2_E =\ell^2 r^2 \Phi \{\Phi^{-2}
(dx^2_0+dx^2_1 + dx^2_2+dx^2_3)
+r^{-4} dr^2+ r^{-2} d\O^2_5\}\;,
\nonumber\\
\Phi&=&(1+a^4r^4)^{1/2}\;, \qquad F_{0
1 2 3 r}=i 4 \ell^{-1} \Phi^{-5/2}
\sqrt{det\;\hat{g}}=i 4 \ell^4 r^3
\Phi^{-4}
\nonumber\\
B_{01}&=&B_{23} =\sqrt{g_s} a^2 r^4
\Phi^{-2}\;, \qquad A_{0 1}=A_{23}
=-i{a^2 \ell^2 \over \sqrt{g_s}} r^4
\Phi^{-2} \;,
\nonumber\\
e^{2\phi}&=&g_s^2 \Phi^{-2}\;, \qquad
\qquad \chi=i {a^4 r^4 \over g_s}\;,
\label{MR}
\eqq
where $g_s$ is the string coupling in
the IR limit $r=0$, and
$a^2=\sqrt{4\pi \a^{'2} g_s
N}/B^{\infty}$, $\ell^2=\a^{'}\sqrt{4
\pi N}$ with $\a^{'}$ the string scale,
$N$ the D3-brane charge and $B^{\infty}$
the boundary value of B-field at $r=\infty$.
Note that the world-volume (specified by
the directions 0,1,2,3) quantities
have been properly re-scaled as in
\cite{MR}, and also that with a self-dual
2-form background (or non-commutativity
parameters) the world volume metric for
the NCYM remains isotropic.

As one can see, the profile of
$\Phi$ is chosen so that the
solutions of NSNS and RR scalars in
(\ref{MR}) satisfy the first
equation of the self-duality
conditions (\ref{B}), which would not
hold for an arbitrary dilaton profile.
Following the proposal of \cite{MR},
we then conclude that the dilatonic
gravity (\ref{can}) with the specific
dilaton radial profile given in
(\ref{MR}) is the holographic dual
of the NCYM with isotropic
non-commutativity. Moreover, we can
read the 5-dimensional background
metrics from (\ref{MR}), that is
\be
\hat{g}_{mn}dx^mdx^n=\ell^2 r^{-2}
\Phi dr^2 + \ell^2 r^2 \Phi^{-1}
dx^2_{\|}
\eq
for the action $I_5$, and
\be
g_{mn}dx^mdx^n=\ell^2 r^{-2}
\Phi^{8/3} dr^2 + \ell^2 r^2
\Phi^{2/3} dx^2_{\|}
\label{metric1}
\eq
for the action $I^{EH}_5$, where
$x_{\|}$ represents the longitudinal
coordinates and $r$ is called the
holographic coordinate which is the
energy scale from the field theory
point of view.

\section{Holographic RG flow of NCYM
in Self-Dual B-background}

The existence of a consistent
effective 5-dimensional dilatonic
gravity allows us to generalize a
counter-term generating algorithm
in the AdS/CFT correspondence, known as
the holographic renormalization group
(RG) flow that is determined by the
dilaton profile\cite{FW}. The dilaton
is interpreted as an effective coupling
running with the energy scale in
the dual field theory. If the dilaton
has a constant radial profile, the theory
reduces to pure AdS gravity with the
holographic dual a CFT having a vanishing
beta function. When the dilaton has a
nontrivial radial profile, the holographic
Callan-Symanzik RG equations have been
constructed by de Boer, Verlinde and
Verlinde in an elegant formalism using the
standard Hamilton-Jacobi theory\cite{BVV,Sa}.
The $c$-functions in the Weyl anomaly can
then be calculated.

The essence of the de Boer-Verlinde-Verlinde
formalism is the observation that though the
equations of motion of the 5-dimensional
supergravity is of second order, the
evolution equation of its on-shell
action $S$, derived from the standard
Hamilton-Jacobi theory, is of first
order and takes the usual form of the
Callan-Symanzik equations, therefore
$S$ can be interpreted as the
4-dimensional quantum effective action
after integrating out all the matter
degrees of freedom coupled to the
background gravity.

According to the holographic interpretation
of the gravity dual, a preferred radial
coordinate in the bulk gravity can be selected
out as representing the energy scale of the
dual field theory. For simplicity, we choose
the "temporal" gauge for 5-dimensional
metric
\be
g_{mn}dx^{m}dx^{n}=d\rho^2+\g_{\sigma
\nu}(\rho,x)dx^{\sigma}dx^{\nu} \;,
\label{metric2}
\eq
where $\rho$ is the holographic radial
coordinate.

For the metric on a boundary screen
located at the radial position $\rho$,
we can further separate out the
holographic coordinate dependence:
\be
\g_{\sigma \nu}(\rho,x) = \mu^2(\rho)
\bar{\g}_{\sigma \nu}(x)\;.
\label{mu}
\eq
where $\bar{\g}_{\sigma \nu}$ is the
background geometry seen by the dual
field theory at some fundamental scale,
and the warp factor $\mu^2$ is the
overall length scale on the boundary
screen. According to the holographic
UV/IR relation\cite{SusW}, $\mu$
stands for the energy scale of
the boundary QFT, and we define
the beta function for the dilaton
$\Phi$ by
\be
\b \equiv \mu {d\Phi \over d\mu}\;,
\label{beta}
\eq
which can be easily calculated once
the 5-dimensional metric and the
dilaton profile are given.

For example, the proposed gravity dual
(\ref{MR}) of NCYM with isotropic
commutativity has the dilaton profile
\be
\Phi=(1+a^4r^4)^{1/2}\;,
\label{phi}
\eq
and the energy scale can be read from
the defining metric (\ref{metric1}),
(\ref{metric2}) and (\ref{mu}):
\be
\mu=\ell\;r\;\Phi^{1/3}={\ell \over
a}\; \Phi^{1/3} (\Phi^2-1)^{1/4}\;,
\label{mu1}
\eq
and the resulting beta function from
(\ref{beta}) and (\ref{mu1}) is
\be
\b={6\Phi(\Phi^2-1) \over 5\Phi^2-2}\;.
\label{betac}
\eq
Note that $\mu$ is a monotonically
increasing function of $\Phi$ and $r$,
so the UV limit $r\rightarrow \infty$
corresponds to $\mu \rightarrow
\infty$ and $\Phi \rightarrow \infty$,
and the IR limit $r \rightarrow 0$ to
$\mu \rightarrow 0$ and $\Phi
\rightarrow 1$.

To develop the Hamilton-Jacobi theory, we
shall cast the 5-dimensional dilatonic
gravity action into the canonical formalism
using the above metric:
\bee
I&=&{1 \over 2 \kappa^2_{5}}
\int_{\cal{M}} d^5x
\sqrt{det\;g}\;[R+ {1 \over 2}
G(\Phi)( \nabla \Phi)^2 +V(\Phi)]
\label{I}
\\
&\equiv& {1 \over 2 \kappa^2_{5}} \int
d\rho \; L\;,
\nonumber\\
L &=& \int d^4x \sqrt{det\;\g}\;
[\pi_{\sigma \nu}\dot{\g}^{\sigma
\nu}+\Pi \dot{\Phi}-\cal{H}]\;.
\eqq
Here $\;\dot{}\;$ denotes the
derivative with respect to $\rho$, and
the canonical momenta and the
Hamiltonian density are defined by
\bee
&& \pi_{\sigma \nu}\equiv {1\over
\sqrt{det\;\g}} {\delta L \over \delta
\dot{\g}^{\sigma \nu}}\;, \qquad \Pi
\equiv {1\over \sqrt{det\;\g}} {\delta
L \over \delta \dot{\Phi}}\;,
\label{momenta}
\\
\cal{H} &\equiv& {1\over
3}\pi^2-\pi_{\sigma \nu}\pi^{\sigma
\nu}+{\Pi^2 \over 2G}-\cal{L}\;,
\\
\cal{L}&\equiv& {\cal{R}}+{1\over
2}\;G\; \g^{\sigma \nu} \del_{\sigma}
\Phi \del_{\nu} \Phi +\;V\;,
\eqq
with $\cal{R}$ the Ricci scalar of the
boundary metric $\g_{\mu \nu}$. Note
that $\int d^4x \cal{L}$ is the action
dimensionally reduced from (\ref{I}).

The defining equations of canonical
momenta (\ref{momenta}) can be
inverted to obtain the first order
flow equations
\bee
\dot{\g}_{\sigma \nu}&=& 2\pi_{\sigma
\nu} - {2 \over 3} \g_{\sigma \nu}
\pi^{\sigma}_{\sigma}\;,
\label{flow1}
\\
\dot{\Phi}&=&{\Pi \over G}\;.
\label{flow2}
\eqq
These equations will be helpful in solving
the resulting Hamilton-Jacobi equation.

In the canonical formulation of the
gravity theory, the Hamiltonian
density $\cal{H}$ gives rise to
a constraint ${\cal H}=0$, imposed
upon the canonical variables:
\be
{1\over 3}\pi^2-\pi_{\sigma
\nu}\pi^{\sigma \nu}+{\Pi^2 \over
2G}={\cal{R}}+{1\over 2}\;G\;
\g^{\sigma \nu} \del_{\sigma} \Phi
\del_{\nu} \Phi +\;V \;.
\label{HJ}
\eq
We then introduce the Hamilton-Jacobi
functional $S$ with a properly assumed
form, and see if we can derive
first-order evolution equations for
terms in $S$. With the hint of the
AdS/CFT correspondence, one interprets
$S$ as the quantum effective action
of the dual field theory after
integrating out the matter degrees
of freedom coupled to the background
gravity, which is assumed of the usual
form on a curved space:
\bee
S[\g,\Phi]&=&S_{EH}[\g,\Phi]+\G[\g,\Phi]\;,
\\
S_{EH}[\g,\Phi]&=& \int d^4x
\sqrt{det\;\g}\;
[Z(\Phi){\cal{R}}+{1\over
2}\;M(\Phi)\g^{\sigma \nu}
\del_{\sigma} \Phi \del_{\nu} \Phi
+U(\Phi)]\;.
\label{S}
\eqq
$S_{EH}$ is the tree level renormalized
action which is similar in structure
to the Lagrangian density $\cal{L}$,
and $\G$ contains the higher-derivative
and non-local terms.

In the Hamilton-Jacobi theory, the
canonical momenta are related to the
Hamilton-Jacobi functional $S$ by
\be
\pi_{\sigma \nu}= {1\over
\sqrt{det\;\g}} {\delta S \over
\delta \g^{\sigma \nu}}\;, \qquad
\Pi = {1\over \sqrt{det\;\g}}
{\delta S \over \delta \Phi}\;.
\label{noether}
\eq
With these relations and the
interpretation of $S$ as the effective
quantum action, the quantum average of
the boundary stress tensor
$<T_{\sigma\nu}>$ and that of the gauge
invariant operator $<O_{\Phi}>$ to which
$\Phi$ couples can be related to $\G$ by
\be
<T_{\sigma \nu}>= {2 \over
\sqrt{det\;\g}} {\delta \G \over \delta
\g^{\sigma \nu}}\;, \qquad <O_{\Phi}>
= {1\over \sqrt{det\;\g}} {\delta \G
\over \delta \Phi}\;.
\eq
The factor of two is determined
from the Hamilton-Jacobi equation
by requiring the correct proportionality
to the beta-function term in the Weyl
anomaly:
\be
<T^{\sigma}_{\sigma}>=\b <O_{\Phi}>-
c\; {\cal{R}}_{\sigma
\nu}{\cal{R}}^{\sigma \nu} +d\;
{\cal{R}}^2\;,
\label{weyl}
\eq
where $\b$ is the beta function
defined in (\ref{beta}), and $c$ and
$d$ are the $c$-functions.

Substituting (\ref{S}) into
(\ref{noether}) we obtain the explicit
form of the canonical momenta, which
will be helpful in solving the
Hamilton-Jacobi equation (\ref{HJ}),
\bee
\pi_{\sigma \nu}&=&{1\over
2}<T_{\sigma \nu}>+Z{\cal{R}}_{\sigma
\nu} +({M \over
2}-Z^{''})\del_{\sigma}\Phi
\del_{\nu}\Phi - Z^{'} \nabla_{\sigma}
\nabla_{\nu} \Phi
\nonumber\\
&-& {1\over 2}\g_{\sigma
\nu}[Z{\cal{R}}+({M\over 2} -2 Z^{''}
)(\nabla \Phi)^2 -2 Z^{'} \nabla^2
\Phi + U] \;,
\label{pi1}
\\
\Pi&=& <O_{\Phi}>+
Z^{'}{\cal{R}}-{M^{'} \over 2} (\nabla
\Phi)^2 - M \nabla^2 \Phi + U^{'}\;,
\label{Phi1}
\eqq
where $\;'\;$ denotes the derivative
with respect to $\Phi$ and the
covariant derivatives here are with
respect to $\g_{\sigma \nu}$.

To derive the evolution equations for
terms in $S_{EH}$, we insert the expansion
(\ref{pi1}) and (\ref{Phi1}) into
(\ref{HJ}), and solve the resulting
Hamilton-Jacobi equation by equating
terms on both sides with the same
functional form. With this procedure
we get from the potential term,
\be
{U^2 \over 3} +{ U^{'2} \over 2 G}=V\;,
\label{U}
\eq
and from the curvature term,
\be
{U \over 3}Z+{U^{'} \over G} Z^{'}=1\;.
\label{Z}
\eq
Note both are first-order evolution
equations.

Moreover, combining the second-order
curvature terms and the first-order
terms in the quantum average
$<T^{\sigma}_{\sigma}>$ and
$<O_{\Phi}>$, we can obtain the
expression for the Weyl anomaly which
is of the form of (\ref{weyl}),
with the $c$-functions given by
\be
c={6Z^2 \over U} \;,\; \qquad
d={2\over U}(Z^2+{3 Z^{'2} \over 2
G})\;.
\label{c}
\eq
We can rewrite the curvature part of
(\ref{weyl}) in terms of the Euler
density $\cal{E}$, the Weyl density
$\cal{W}$ and the Ricci scalar squared
as follows
\be
c\; {\cal{R}}_{\sigma
\nu}{\cal{R}}^{\sigma \nu} +d\;
{\cal{R}}^2 = {c \over
2}\;({\cal{E}}-{\cal{W}})+ (d-{c\over
3})\;{\cal{R}}^2\;,
\eq
where
\be
{\cal{E}}={\cal{R}}^2-4\;{\cal{R}}_{\sigma
\nu}{\cal{R}}^{\sigma
\nu}+{\cal{R}}_{\sigma \nu \lam
\delta}{\cal{R}}^{\sigma \nu \lam
\delta}\;, \qquad {\cal{W}}={1\over
3}{\cal{R}}^2-2\;{\cal{R}}_{\sigma
\nu}{\cal{R}}^{\sigma
\nu}+{\cal{R}}_{\sigma \nu \lam
\delta}{\cal{R}}^{\sigma \nu \lam
\delta}\;.
\eq
Note that $\cal{E}$ is a topological
density, and $\cal{W}$ is an invariant
under Weyl transformations, so that the
combination ${\cal{E}}-{\cal{W}}$ is
invariant up to a total derivative
under Weyl transformations. However,
the ${\cal{R}}^2$ term is not a Weyl
invariant, whose presence signals the
non-conformal nature of NCYM when
$c\ne 3d$, as shown later.

Because of the nonlinearity, it is
difficult to solve $U$ from (\ref{U}).
We, however, can solve it from the
flow equations by substituting
(\ref{pi1}) and (\ref{Phi1}) into
(\ref{flow1}) and (\ref{flow2}).
Assuming that the theory is at
sufficiently low energy scale compared
to the cutoff so that the potential
term dominates, it then yields
\bee
U&=&{ 6\; \dot{\mu} \over \mu}\;,
\label{U1}
\\
\b &\equiv& \mu\;{d\Phi \over d\mu}={6
U^{'} \over G U}\;.
\label{U2}
\eqq
Clearly the effective cosmological
constant $U$ is over-determined by
three equations (\ref{U}), (\ref{U1})
and (\ref{U2}), the consistency of the
solutions among them will imply the
validity of the formalism and the
assumption of potential dominance,
which reminds us that the theory is at
sufficiently low energy scale.

On the other hand, the effective
inverse Newton constant $Z$  will be
determined by (\ref{Z}) up to an
integration constant given by the
initial conditions. There are also
equations determining $M$ in the
kinetic term from the input of $U$ and
$Z$; however, we omit them since our
interest is the $c$-functions which are
independent of $M$, and it is easy to show
that $M$ can be consistently solved from
the Hamilton-Jacobi equation.

Having the formalism at hand, we are
ready to calculate the running
behavior of the quantum effective
action $S$ for the NCYM from its dilatonic
gravity dual defined by (\ref{can})
and (\ref{metric1}). The beta function
for $\Phi$ has been given in
(\ref{betac}). Compare (\ref{can}) and
(\ref{I}), we have
\be
G(\Phi)= -20\Phi^{-2}/3\;, \; \qquad
V(\Phi) = \ell^{-2}\Phi^{-8/3}(20-8
\Phi^{-4})\;.
\label{GV}
\eq
With these data and eq. (\ref{betac})
for the beta function, we find the
solutions for the effective cosmological
constant $U$ from the three equations
mentioned above agree with each other,
all giving
\be
U={2 \Phi^{-10/3}\over \ell}(5\Phi^2-2)\;.
\eq

The running behavior of the effective
inverse Newton constant is then
determined from (\ref{Z}) and is given by
\be
Z={\ell \over 6}
\;\Phi^{-2/3}(\Phi^2+2) + Z_0 \;
\Phi^{-2/3} (\Phi^2-1)^{-1/2} \;.
\eq
Note that the second term blows up
in the IR limit $\Phi \rightarrow 1$
if $Z_0 \ne 0$, which will violate
the assumption of potential dominance
at low energy scale; and thus we are
forced to set $Z_0=0$.

Finally, from (\ref{c}) the resulting
$c$-functions are (for $Z_0=0$)
\be
c={\ell^3 \over
12}\;{\Phi^2(\Phi^2+2)^2 \over
5\Phi^2-2}\;, \qquad d={\ell^3 \over
60}\;{\Phi^2(\Phi^4+8\Phi^2+6) \over
5\Phi^2-2}\;.
\label{cfun}
\eq
In the above equations (\ref{GV}) to
(\ref{cfun}), the profile of the dilaton
$\Phi$ is given by eq. (\ref{phi}). Like
the beta function, the $c$-functions
are monotonically increasing with
$\Phi$ (and thus with $\mu$) for $\Phi
\ge 1$. This is a generalization of
Zamolodchikov's C-theorem\cite{c-thrm}
in two dimensions that the c
functions are always monotonically
increasing with the energy scale; so
one may say that {\it{the C-theorem
holds true in the present case}}. Away
from the IR limit, $\Phi>1$ and the ratio
$c/d\neq 3$, differing from the one
($c/d=3$) for ordinary Yang-Mills theory
in the usual AdS/CFT correspondence with
$\Phi\equiv 1$, which is the IR limit
of the NCYM.

\section{$c$-functions as vectors on
the $\Phi$-Space}
In the section II we have seen that
the form of the 5-dimensional dilatonic
gravity action, dimensionally reduced
from 10-dimensional supergravity as
the dual of NCYM, is not unique. We have
obtained two such actions, one given by
(\ref{can}) with a canonical
Einstein-Hilbert term for gravity and
the other (\ref{noncan}) of a
non-canonical form; they are related to
each other by a Weyl transformation
(\ref{weylt}). In the section III, we
have chosen to work with the canonical
form of the action (\ref{can}). One
may wonder what are the resulting
beta and $c$-functions if we work with
the non-canonical action (\ref{noncan}).
The answer for the beta function is
straightforward: from its definition
(\ref{beta}), it should transform as
a vector on the $\Phi$-space which can
be thought as the coupling constant
space of NCYM. To be explicit, let
us call the energy scale parameter
$\mu_q$\footnote{The subscript $q$
will be used to specify the
non-canonical counterparts of the
quantities defined in section III;
the same convention will be also
adopted in the Appendix.} for
the non-canonical gravity in contrast
to the parameter $\mu$ defined for the
canonical one. These two quantities
are related to each other by the Weyl
transformation (\ref{weylt}), which
through (\ref{mu}) leads to
\be
\mu_q =\Phi^{-5/6} \mu \;.
\label{newmu}
\eq
From this, the beta functions in the
two cases are related by
\bee
\b &\equiv& \mu {d\Phi \over d\mu} =
\O \mu_q {d\Phi \over d\mu_q}
\equiv \O \b_q \;,
\\
\O &\equiv& {\mu \over \mu_q}\;
{d\mu_q \over d\mu}={3 \over
5\Phi^2-2}\;.
\eqq

Though the beta function has clear
geometric meaning by its definition,
it is not clear if the $c$-functions
have also the geometric meaning as
vectors on the $\Phi$-space.
To answer this question, we need to
generalize the de-Boer-Verlinde-Verlinde
formalism to the non-canonical action.
The generalization is straightforward
but tedious, we will leave the details
to Appendix A. The resulting $c$-
functions turn out to be
\be
c_q =\O^{-1} c \;, \qquad d_q =\O^{-1}
d\;,
\label{geom}
\eq
and are thus vectors on the
$\Phi$-space. Note that
(\ref{geom}) is true as long as the
integration constants for $Z_q$ and
$Z$ are set to equal, that is
\be
Z_q=\Phi^{5/3} Z={\ell \over 6}
\;\Phi(\Phi^2+2) + {Z_0 \Phi \over
\sqrt{\Phi^2-1}}  \;.
\eq

Now that the $c$-functions have a
geometric interpretation, it would be
interesting to see if the C-theorem
may have a generic geometric origin.
This issue has been explored in the
recent works\cite{C-geom} on the
gravity side. We leave this problem
for NCYM for future study.

\section{Conclusions and Discussions}
Since Maldacena and Russo proposed the
supergravity dual of NCYM, not much
has been done along this line. In this
paper, we first pointed out that the
gravity dual of NCYM with isotropic
non-commutativity has a consistent
5-dimensional action in the form of a
dilatonic gravity, which enables us
to adopt the holographic RG flow
approach to investigate the physics on
the dual field theory side,
generalizing the usual AdS/CFT
correspondence.

We adopted the de Boer-Verlinde-Verlinde
formalism to evaluate the $c$-functions
at low energies, under the assumption of
potential dominance, and found that the
$C$-theorem holds true in the present case.
The ratio of the two  coefficient fucntions
in the Weyl anomaly away from the IR limit
is different from that in ordinary Yang-Mills
theory, indicating the non-conformal nature
of the NCYM. All of these were seen from
the dual gravity side. To examine these
phenomena directly inside the NCYM itself
is worthwhile, especially because the
perturbative techniques of non-commutative
field theory seem to have become matured
in a series of recent works\cite{SR,PNC}.

The calculations of Weyl anomaly and
boundary counter-terms for the
boundary conformal theory from the AdS
gravity have been performed in many
different ways\cite{HS,Bala}, they all
agree to each other. Not much similar
efforts have been spent for the
non-commutative cases. Besides the
method adpoted in this paper, there is
an alternative approach\cite{Odin} by
generalizing the method of Henningson
and Skenderis\cite{HS} to dilaton
gravity which applies only to
asymptotically AdS spacetime. However,
as pointed out in the second paper of
ref. \cite{Odin}, the NCYM dual at
hand, corresponding to eq. (57) there,
has not asymptotic AdS region in UV. It
would be interesting to see whether an
improvement of their approach can be
applied to the NCYM dual.

Although we have defined the
$c$-functions from its gravity dual by
calculating the Weyl anomaly, we still
lack a general understanding from the
field theory side. It has been an issue
of defining sensible $c$-functions in
4 dimensions and there is an on-going
debate about the validity of a general
4-dimensional C-theorem\cite{Cardy}.
In section IV, we clarified the nature
of $c$-functions on the coupling constant
space, and showed they are indeed vectors
on the coupling constant space as the
beta function. We hope this geometric
understanding will help in
constructing a geometric realization
of the C-theorem in 4 dimensions.

Up to now, we have only considered the
supergravity background with self-dual
B-field configurations. It would be
interesting to consider more general
B-backgrounds, which will correspond to
NCYM with anisotropic non-commutativity.
The 5-dimensional gravity dual will
then be a dilaton gravity coupled to
the 2-form potentials, and we need to
generalize the de Boer-Verlinde-Verlinde
formalism to include the dynamics of
2-form potentials, which may help us to
understand more about the physics of
NCYM from its gravity dual.

\renewcommand{\theequation}{A.\arabic{equation}}
\setcounter{equation}{0}

\bigskip

\noindent
\appendix
\section{Generalization of de
Boer-Verlinde-Verlinde
Formalism For Non-Canonical Gravity }

We start with the non-canonical action
(\ref{noncan}) and cast it into the
ADM form as done for the canonical one:
\bee
I_5&=&{V_5 \over 2 \kappa^2_{10}} \int
d^5x \sqrt{det \; \hat{g}} \;
[X_q(\Phi)\hat{R}+{1\over 2} G_q(\Phi)
(\hat{\nabla} \Phi)^2 + V_q(\Phi)]
\\
&\equiv& {V_5 \over 2
\kappa^2_{10}}\int dr L_q \;,
\eqq
where $X_q$, $G_q$ and $V_q$ can be
read from (\ref{noncan}).

Decompose the metric into the warped
form
\be
\hat{g}_{mn}dx^{m}dx^{n}=N^2d\rho^2+\g_{\sigma
\nu}(\rho,x)dx^{\sigma}dx^{\nu} \;,
\qquad N=\pm 1\;,
\eq
(with $N$ the lapse function). Using
the identity
\be
\hat{R}=-2\nabla_{m}(n^m \nabla_{n}n^n)
+{\cal{R}}-({\cal{K}}_{\sigma \nu }
{\cal{K}}^{\sigma \nu }- {\cal{K}}^2)
\eq
where $n^m$ is the boundary unit
normal, $\cal{R}$ and $\cal{K}$ are
the intrinsic and extrinsic boundary
curvature respectively, we then have
\be
L_q=\int d^4x \sqrt{det \;
\g}\;[\pi_{\sigma \nu}\dot{\g}^{\sigma
\nu}+\Pi \dot{\Phi}-N \cal{H}]\;,
\eq
with
\bee
\cal{H} &\equiv& N[{1\over
X_q}({1\over 3}\pi^2-\pi_{\sigma
\nu}\pi^{\sigma \nu})+{1\over
2G_q}\Pi^2+ {X^{'}_q \pi \dot{\Phi}
\over 3 X_q}-{X^{'}_q {\cal{K}}\;\Pi
\over G_q N} -\cal{L}\;]\;,
\label{hjq}
\\
\cal{L}&\equiv& X_q \;
{\cal{R}}+{1\over 2}\;G_q \; \g^{\sigma
\nu} \del_{\sigma} \Phi \del_{\nu}
\Phi +V_q\;,
\eqq
where $\;'\;$ denotes derivative with
respect to $\Phi$, and $\; \dot{} \;$
with respect to $\rho$.

The canonical momenta are defined by
\bee
\pi_{\sigma \nu}&\equiv& {1\over
\sqrt{det\;\g}} {\delta L_q \over
\delta \dot{\g}^{\sigma \nu}}={X_q
\over N}\;({\cal{K}}_{\sigma
\nu}-\g_{\sigma \nu}
{\cal{K}})-\g_{\sigma \nu} X^{'}_q
\dot{\Phi}\;,
\\
\Pi &\equiv& {1\over \sqrt{det\;\g}}
{\delta L_q \over \delta
\dot{\Phi}}=G_q \dot{\Phi}+{2 X^{'}_q
{\cal{K}} \over N}\;.
\eqq
By inverting these equations, we
obtain the flow equations
\bee
{\cal{K}}_{\sigma \nu}&\equiv& {1\over
2 N} \dot{\g}_{\sigma \nu}={N \over
X_q}[\pi_{\sigma \nu} -{1\over 3}
\g_{\sigma \nu} \;(\pi + X^{'}_q
\dot{\Phi})]\;,
\\
\dot{\Phi}&=& F \;(\Pi+ {2 X^{'}_q
\over 3 X_q} \; \pi) \;, \qquad F
\equiv {1 \over G_q-{8X^{'2}_q \over
3X_q}}  \;,
\eqq
and then substituting these two equation
into (\ref{hjq}), the Hamiltonian
density $\cal{H}$ can be expressed
completely in terms of the canonical
momenta.

Define the Hamiltonian-Jacobi
functional as before
\be
S[\g,\Phi]=\G[\g,\Phi]+ \int dx^4
\sqrt{det\;\g}\;
[Z_q(\Phi){\cal{R}}+{1\over
2}\;M_q(\Phi) (\nabla \Phi)^2
+U_q(\Phi)]\;,
\eq
and solve the Hamiltonian-Jacobi
equation and the flow equations by
adopting the new energy scale
parameter $\mu_q$ defined in
(\ref{newmu}). We obtain the beta
function
\be
\beta_q
=\O^{-1}\beta=2\Phi(\Phi^2-1)\;,
\qquad
\O \equiv {3 \over 5\Phi^2-2} \;,
\eq
and the renormalized dilatonic
potential and coefficient of the
scalar curvature
\be
U_q=\Phi^{10/3}U={10\Phi^2-4 \over
\ell}\;, \qquad Z_q=\Phi (\Phi^2+2) +
{Z_{q0} \Phi \over \sqrt{\Phi^2-1}}\;.
\eq
As mentioned in section IV, if we take
$Z_{q0}=Z_0$, then $Z_q=\Phi^{5/3}Z$,
and the resulting $c$-functions
transform as vectors on the $\Phi$-space.

The formal expressions for the
$c$-functions are somewhat involved:
\bee
c&=&{1\over T}\;{Z_q^2 \over X_q}\;,
\\
d&=&{1\over T}\;[{Z_q^2\over 3X_q} +
{Z^{'2}_q \over 2 G_q} - Z_q H
-{X^{'}_q Z^{'}_q Z_q \over 3 G_q X_q}
+ {4 X^{'}_q Z^{'}_q H \over G_q
}\;]\;,
\eqq
with
\bee
T&\equiv& {-U_q \over 3 X_q} + {F
X^{'}_q \over 3 X_q}(U^{'}_q -{8
X^{'}_q U_q \over 3 X_q})+ {X^{'}_q
U^{'}_q \over 3 G_q X_q} +{8 F
X^{'3}_q U^{'}_q \over 9 G_q X^2_q} \;,
\\
H&\equiv&{F X^{'}_q (Z^{'}_q -{2
X^{'}_q Z_q \over 3 X_q})\over 3 X_q}
\;.
\eqq
However, the final expressions are
very simple
\be
c_q =\O^{-1} c \;, \qquad d_q =\O^{-1}
d\;,
\eq
as long as $Z_{q0}=Z_0$. Indeed, by
continuity at $\Phi=1$, we are forced
to take $Z_{q0}=Z_0=0$.

\end{document}